# LOW-COMPLEXITY VIDEO ENCODER FOR SMART EYES BASED ON UNDERDETERMINED BLIND SIGNAL SEPARATION

*Jing Liu, Fei Qiao*, Zhijian Ou and Huazhong Yang*

Department of Electronic Engineering, Tsinghua University

## ABSTRACT

This paper presents a low complexity video coding method based on Underdetermined Blind Signal Separation (UBSS). The detailed coding framework is designed. Three key techniques are proposed to enhance the compression ratio and the quality of the decoded frames. The experiments validate that the proposed method costs 30ms encoding time less than DISCOVER. The simulation shows that this new method can save 50% energy compared with H.264.

## 1. INTRODUCTION

Recently, the smart camera systems gained more popularity with the emergence of wireless video surveillance, multimedia sensor network and image capturing of mobile devices [1]. However, limited power supply and low computational ability of the encoder impedes the wide implementation of such novel applications with a sufficient image/video compression ratio. Therefore, low-complexity video encoders with less computation complexity and power consumption are necessary to cater for such given smart eyes.

Multiple attempts in three key categories have been made to reduce the encoding computation complexity. The first one is the improvement of the conventional video encoder including simplify motion estimation, DIC/IDCT and quantization which almost occupy 93% encoding computation time [2]. But the computation complexity is still high after the enhancement. The second one is the Distributed Video Coding (DVC) approach developed based on Slepian-Wolf (SW) theory and Wyner-Ziv (WZ) theory [3, 4]. Compared to conventional video codec, the complexity distribution of DVC is swapped. But its decoder is too complex to decode the video frames in real time. Moreover, Compressive Sensing (CS) method is used to compress video sequence with low complexity consumption [5, 6], but it is very difficult to trade off between compression ratio and recovered picture quality. So far, there are no acceptable standards for the low-complexity encoding applications. New efforts are consistently being made to improve existing low-complexity encoding methods or develop new ones.

Motivated by the fact that several sounds can be picked out from a mixed audio signal in a noisy cocktail party, we propose a novel video compression approach that several video frames are first encoded as one mixed frame. At the decoder side, the mixed frame is separated into the estimations of the previous video frames via Underdetermined Blind Signal Separation (UBSS). The UBSS framework is a natural fit for low-complexity video encoding, because the involved computation of the encoder side is only the matrix multiplication.

This paper is organized as follows. The section II briefly reviews the related BSS problem. In section III, the detailed structure of the new approach is first provided, and three key technologies are introduced. And section IV shows simulation result to validate this method. Finally, section V summarizes the proposed method

## 2. ENCODING AND DECODING ALGORITHM

Blind Signal Separation was first established by J. Herault and C. Jutten in 1985 [7], aimed at recovering the unknown source signals only by several observed linear mixed sources. It can be described by the following equations.

$$y = Wx = WAs \qquad (1)$$

where $s(t) = [s_1(t) \quad s_2(t) \quad \cdots \quad s_n(t)]$ is the unknown source signals matrix. $x(t) = [x_1(t) \quad x_2(t) \quad \cdots \quad x_m(t)]$ is the observed signals matrix. $y$ is the estimation of $s$. $A \in R^{m \times n}$ is the mixing matrix. $W \in R^{n \times m}$ is the separating matrix. In this paper, the randomly Gaussian matrix is used as the mixing matrix [8]. The consecutive video frames $f_1, f_2, \cdots, f_n$ are taken as the source signals $s$ in equation (1) and disregarded their time sequence. Varying from the traditional UBSS problem, the order of the recovered frames would not be disrupted because the mixing matrix is known exactly to the separation side. The coding process is shown in Fig. 1.

At present, there are three types of separation algorithms: greedy algorithm, ℓ1 minimization, Total Variation (TV) minimization. TV minimization can recovery those signals or images which are not sparse but its gradient is sparse [9], which is suitable for the video frames. In order to improve TV minimization's computation complexity and guarantee its robustness, C. Li introduced the augmented lagrangian method to solve TV minimization problem and proposed TVAL3 algorithm in 2009 [9]. In our paper, TVAL3 is adopted as the separation algorithm because it has better performance than greedy algorithm, ℓ1 minimization and other TV minimization algorithms. Of course, there are inherent errors of TVAL3 [9]. Also, the size of source signals matrix influences the quality of recovered frames and separation time [9].

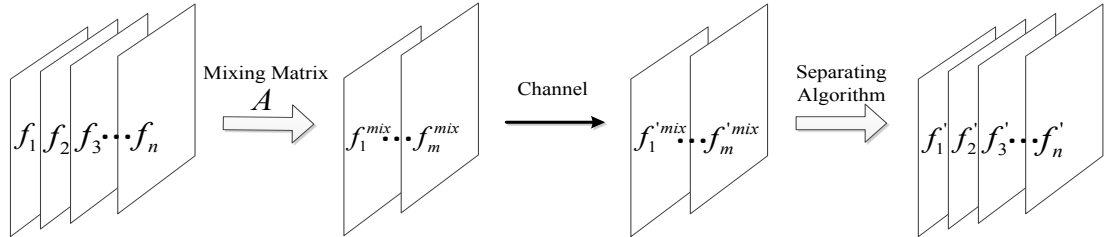

Fig. 1 Video coding process by UBSS (n>m)

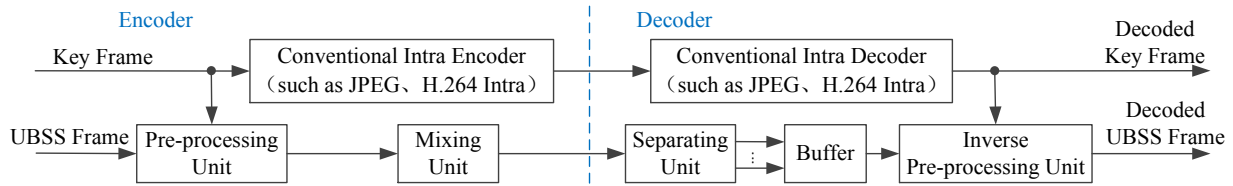

Fig. 2 Video compression framework based on UBSS

## 3. CODER IMPLEMENTATION

### 3.1 Codec Structure

For Underdetermined Blind Signal Separation (UBSS), the amount of source signals is more than that of observed signals. So it can be properly used to compress the video sequence[10]. The proposed video compression framework based on UBSS is shown in Fig. 2. The video frames are divided into key frames and UBSS frames. The key frames are encoded and decoded by conventional intra coding methods such as JPEG, H.264 Intra. While the UBSS frames are encoded by underdetermined mixing and decoded by separation algorithm of UBSS. At the encoder side, there is a pre-processing unit, which usually needs the information of key frame, before mixing unit.

For the proposed low-complexity video encoder, three technologies are employed to enhance the decoding quality, lower the decoding complexity and reduce the memory consumption for encoding. First, for the pre-processing unit shown in Fig. 2, the key technology that the UBSS frames first subtract their preceding key frames before mixing is used to ensure the recovered quality. Second, for the mixing unit, there are two key technologies. One is the streamed-way mixing approach to save memory at the encoder side. The other is the block-level mixing method for decoding complexity reduction and encoding quality enhancement. A detailed explanation is provided as follows.

### 3.2 Residual Mixing

In this work, the pre-processing unit for UBSS frame in Fig. 2 is to subtract the key frame. Only the residuals are mixed. We call this approach as "Residual Mixing". There are two advantages by performing Residual Mixing. Firstly, the decoding quality will be better due to the sparser residuals' gradient. So the compression ratio of UBSS frame could be increased. Secondly, by subtracting the preceding key frame, the relative error of decoded UBSS frame caused by inherent errors of TVAL3 can be reduced. At the decoder side, the UBSS frame residuals are first decoded by TVAL3. And then, these recovered residuals are added to the relevant key frame to get the decoded UBSS frames.

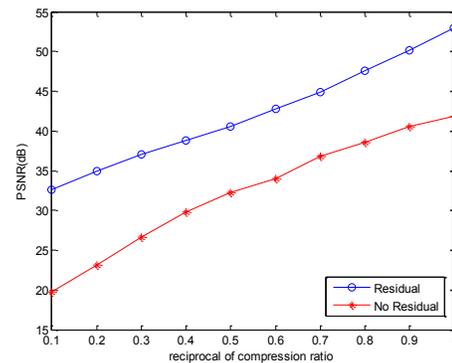

Fig. 3 Effect of residual mixing on separation quality

To validate the effect of this approach, experiments are performed on video sequence "Foreman", whose spatial resolution is QICF (144*176). The separation effect is evaluated by Peak-Signal-to-Noise Ratio (PSNR). Fig. 3 shows the experiment result. The x axis represents the reciprocal of compression ratio. It is obvious that the quality of recovered image after subtracting key frame is 10dB better than that of non-residual mixing.

### 3.3 Streamed-Way Mixing

The UBSS frames are encoded by mixing several UBSS frames. If the mixing process is conducted after all needed UBSS frames are sampled and stored at the encoder side, it requires a lot of memory consumption, shown in Fig. 4(a). Here we take n=4 for example.

In order to save the memory consumption, special mixing approach is needed. In this application, mixing matrix is fixed, it is possible that the frames are encoded in sampling order as shown in equation (2), where $a_i$ is the i-th column of $A$. Therefore, the streamed-way mixing can be achieved by dividing $A$ into several subsets $A_i^{'}$, which is composed by the i-th column of mixing matrix $A$, shown in Fig. 4(b).

$$x = A[f_1 \ f_2 \ f_3 \ f_4]^T = a_1 f_1 + a_2 f_2 + a_3 f_3 + a_4 f_4 \\ = A_1^{'} f_1 + A_2^{'} f_2 + A_3^{'} f_3 + A_4^{'} f_4 \quad (2)$$

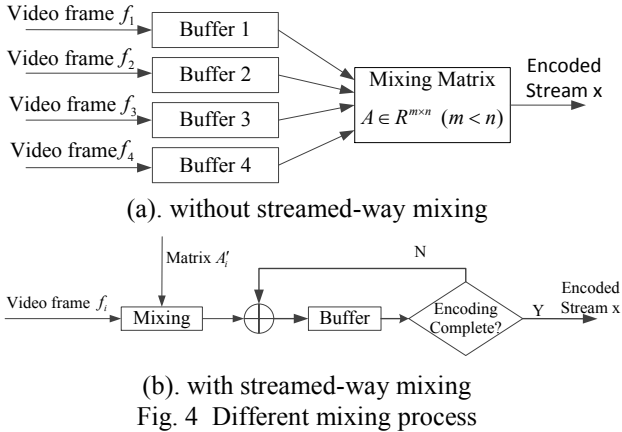

(a). without streamed-way mixing

(b). with streamed-way mixing

Fig. 4 Different mixing process

By mixing in a streamed-way, the memory size needed in the encoder side for storing frames is equal to the final compressed frame size. Compared (a) and (b), it is obvious that mixing in a streamed-way can save lots of memories.

### 3.4 Block-Level Mixing

In this work, UBSS frames are being mixed at block-level. This is because that for TVAL3 algorithm, when the compression ratio is fixed, the input size not only influences the quality of recovered image, but also affects decoding time. To validate this conclusion, multiple experiments are conducted on the same video sequence "Foreman". Table I shows the results.

The result shows that when both decoding time and PSNR are taken into account, the optimum block size is 32×32. In our simulation, the 32×32 block is composed by 4 corresponding 16×16 blocks from 4 consecutive UBSS frames shown in Fig. 5. Moreover, it is very efficient to employ the temporal correlation among frames by choosing corresponding blocks from consecutive UBSS frames.

Table I. Decoding time and PSNR for different block size

| Block Size | 4*4 | 8*8 | 16*16 | 32*32 | 64*64 |
|---|---|---|---|---|---|
| Time(s) | 78.842 | 24.711 | 12.355 | 14.976 | 43.103 |
| PSNR(dB) | 8.920 | 21.104 | 21.841 | 29.975 | 33.375 |

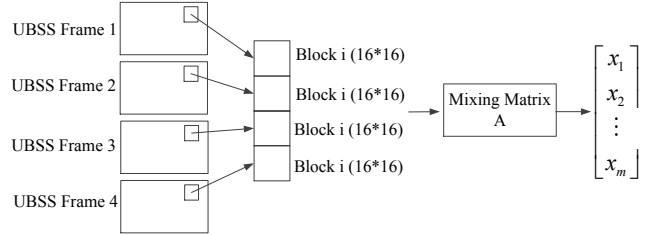

Fig. 5 Mixing process in block-level

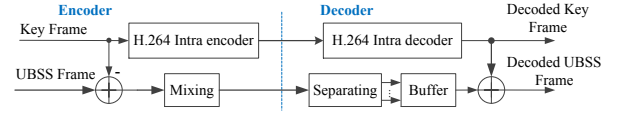

Fig. 6 Detailed codec structure of UBSS+H.264

## 4. SIMULATION RESULTS AND ANALYSIS

This section compares the performance of the proposed method and three well-known low-complexity video coding methods, H.264 Intra, H.264 No motion and DISCOVER. H.264 Intra represents that each video frame is encoded by Intra coding method. H.264 No motion means that the motion vector for motion estimation is set to zero. DISCOVER is one of the most well-known low complexity video coding methods based on DVC. Fig. 6 shows the detailed structure adopted for performance comparison, defined as UBSS+H.264. In this structure, H.264 Intra is used as the key frame encoding method. And three key technologies are adopted as well. The compression performance, encoding time, hardware resource consumption are chosen as there criteria to evaluate the performance of four different video coding approaches.

### 4.1 Compression Performance

Compression performance is one key performance criterion of a video compression standard. It represents the relationship between the quality of recovered image and the compression ratio. For an excellent video coding method, its recovered image's quality is higher than others' at the same compression ratio.

Experiments are conducted on "Foreman". The experiment results are shown in Fig. 7.

The results show that with the increment of compression ratio, PSNR of UBSS+H.264 decreases slower than that of the others. When compression ratio is small, PSNR of UBSS+H.264 is the lowest. It is caused by the error of TVAL3 algorithm. But when compression ratio is larger than 30, UBSS+H.264's PSNR exceeds H.264 Intra's PSNR and is comparable with H.264 No Motion and DISCOVER.

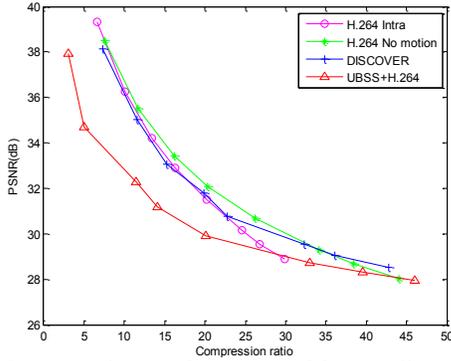

Fig. 7 Compression performance of four coding methods

### 4.2 Encoding time

The encoding complexity of the proposed video codec is composed by two parts, the encoding complexity of the key frames and that of the UBSS frames. There are many ways to measure the encoding complexity. But most of them are very difficult to implement. So in this paper, encoding time is used as a measurement of encoding complexity.

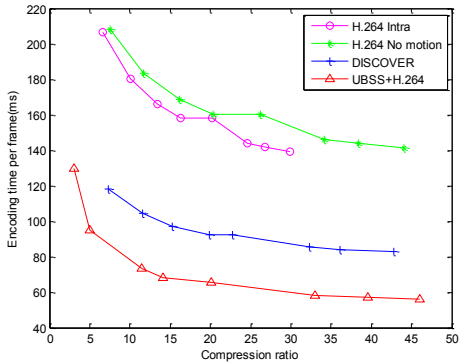

Fig. 8 Encoding time of four coding methods

The test is conducted on the computer with Intel i5 CPU at 2.67GHz, 2GB RAM and 32-bit Win7 operative system. The simulations are done with C++ code using release mode of Visual Studio 2008. In order to ensure that the simulation results are not influenced, nothing is run on the test PC expect operative system. The experiments are still conducted on "Foreman". Fig. 8 shows the experiment result.

The results demonstrate that UBSS+H.264 have the lowest encoding time. The proposed video codec consumes 60%, 57%, 30% lower encoding time than H.264 No motion, H.264 Intra and DISCOVER respectively at the point of compression ratio 30.

### 4.3 Hardware Resource Consumption

In this section, energy consumption and the number of equivalent gate are used to measure the complexity of the proposed coding method. The energy consumption per UBSS frame is calculated by the following steps. First, the power consumption of UBSS frames is simulated by Design Compiler (DC), then multiply the power consumption by the encoding times per UBSS frame to get the energy consumption per UBSS frame shown in table II.

The results show that with the increment of compression ratio, the energy consumption per UBSS frame decreases respectively, an important advantage of the proposed method. This is because that the multiplication times needed is reduced with the reduction of mixing matrix size. Also, the number of equivalent gate for UBSS encoding frame is 980, which is significantly less than that of H.264 Intra. So the proposed framework UBSS+H.264 can efficiently enhance the compression performance of H.264 Intra only at little expense of complexity. Fig. 9 shows the original frame, the decoded results of H.264 and the proposed method. Compared with the decoded result of H.264, the decoded frames' quality of proposed method is acceptable.

Table II. The energy consumption per UBSS frame (mJ)

| 1/Compression Ratio | 0.02 | 0.05 | 0.08 | 0.10 | 0.50 |
|---|---|---|---|---|---|
| Energy(mJ) | 0.007 | 0.016 | 0.026 | 0.032 | 0.159 |

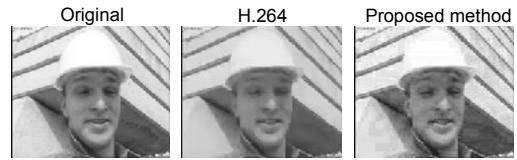

Fig. 9 Comparison results of original frame, H.264 and proposed method decoded frames

### 5. CONCLUSION

In this paper, a new low-complexity video coding method based on UBSS is proposed. The codec structure is presented. Three key technologies--residual mixing, streamed-way mixing, block-level mixing--are proposed to enhance the compression performance. The experiment results demonstrate that the proposed method can achieve comparable compression ratio at very low expense of encoding time. What's more, the proposed approach has an important advantage that the energy consumption per UBSS frame decreases with the increment of compression ratio.

# 6. REFERENCES

[1]. X. Kunzhi, Q. Fei, Q. Wei, and Y. Huazhong, "Smart-Eyes: a FPGA-based smart camera platform with efficient multi-port memory controller," in ICMT-13, 2013.

[2]. Y. K., J. Lv, J. Li, and S. Li, "Practical real-time video codec for mobile devices," in ICME '03, pp. III-509-12 vol.3.

[3]. B. Girod, A. M. Aaron, S. Rane, and D. Rebollo-Monedero, "Distributed Video Coding," Proceedings of the IEEE, vol. 93, pp. 71-83, 2005.

[4]. F. Dufaux, W. Gao, S. Tubaro, and A. Vetro, "Distributed Video Coding: Trends and Perspectives," Eurasip Journal on Image and Video Processing, 2009.

[5]. M. Wakin, J. Laska, M. Duarte, D. Baron, S. Sarvotham, D. Takhar, K. F. Kelly, and R. G. Baraniuk, "Compressive imaging for video representation and coding," in Picture Coding Symposium, 2006.

[6]. M. F. Duarte, M. A. Davenport, D. Takhar, J. N. Laska, T. Sun, K. F. Kelly, and R. G. Baraniuk, "Single-pixel imaging via compressive sampling," Signal Processing Magazine, IEEE, vol. 25, pp. 83-91, 2008.

[7]. A. Cichocki and S. i. Amari, Adaptive blind signal and image processing: learning algorithms and applications. Chichester, England ; New York: John Wiley, 2002.

[8]. R. G. Baraniuk, "Compressive sensing [lecture notes]," Signal Processing Magazine, IEEE, vol. 24, pp. 118-121, 2007.

[9]. C. Li, "An efficient algorithm for total variation regularization with applications to the single pixel camera and compressive sensing," Rice University, 2009.

[10]. J. Liu, F. Qiao, Q. Wei, and H. Yang, "A Novel Video Compression Method Based on Underdetermined Blind Source Separation," in Multimedia and Ubiquitous Engineering. vol. 240, ed: Springer Netherlands, 2013, pp. 13-20.